# A Study on Enhancing User Engagement by Employing Gamified Recommender Systems

Ali Fallahi Rahmatabadi[a], Azam Bastanfard[a]*, Amineh Amini[a], Hadi Saboohi[a]

[a] *Department of Computer Engineering, Karaj Branch, Islamic Azad University, Karaj, Iran*

ali.fallahi@kiau.ac.ir, bastanfard@kiau.ac.ir*, aamini@kiau.ac.ir, saboohi@kiau.ac.ir

## Abstract

Providing customized products and services in the modern business world is one of the most efficient solutions to improve users' experience and their engagements with the industries. To aim, recommender systems, by producing personalized recommendations, have a crucial role in the digital age. As a consequence of modern improvements in the internet and online-based technologies, using gamification rules also increased in various fields. Recent studies showed that considering gamification concepts in implementing recommendation systems not only can become helpful to overcome the cold start and lack of sufficient data, moreover, can effectively improve user engagement. Gamification can motivate individuals to have more activities on the system; these interactions are valuable resources of data for recommender engines. Unlike the past related works about using gamified recommendation systems in different environments or studies that particularly surveyed gamification strategies or recommenders separately, this work provides a comprehensive review of how gamified recommender systems can enhance user engagement in various domain applications. Furthermore, comparing different approaches for building recommender systems is followed by in-depth surveying about investigating the gamified recommender systems, including their approaches, limitations, evaluation metrics, proposed achievements, datasets, domain areas, and their recommendation techniques. This exhaustive analysis provides a detailed picture of the topic's popularity, gaps, and unexplored regions. It is envisaged that the proposed research and introduced possible future directions would serve as a stepping stone for researchers interested in using gamified recommender systems for user satisfaction and engagement.

## 1. Introduction

Since the past decade, due to the swift development of the internet, we have experienced many fundamental changes in different aspects of our life. This new era, also known as the digital age, provided unique opportunities to have more effective communication and engagement with users (Hassan & Hamari, 2020; Nguyen et al., 2018). Companies can use these new digital interaction channels to enhance their transparency and improve their financial benefits (Davidaviciene et al., 2017; Kartajaya et al., 2019). Telling exciting stories about a brand's history, describing the firm's perspectives and future achievements are other potentials of new digital media to increase user's attraction (Fombrun, 1996; Van Riel & Fombrun, 2007).

Increasing users' engagement and satisfaction by concerning their expectations and providing personalized suggestions through deeply customized channels (Camilleri, 2015) Based on this concept, recommender systems help users access what they need quickly and efficiently (Abdollahpouri et al., 2020). *Figure 1* shows a sample scenario of how recommender systems by providing a deeply personalized communication channel for suggesting customized recommendations can increase the user's engagement.

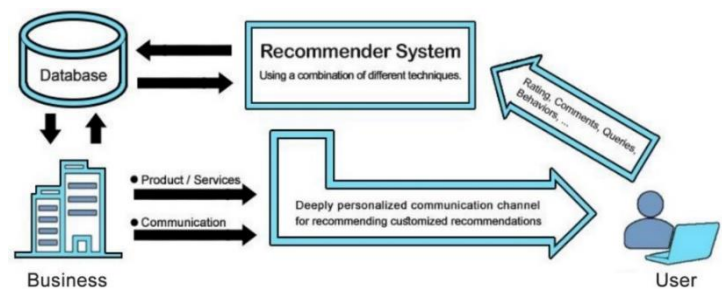

**Figure 1.** A sample scenario of how a recommender system can enhance communication and user satisfaction

Another digital age accomplishment is gamification. Due to the recent developments in internet-based technologies, using gamification practices has also been raised in many domains such as virtual organizations (Rapp, 2020), tech companies (Xi & Hamari, 2019), marketing (Lucassen & Jansen, 2014), business applications (Conaway & Garay, 2014), online learning platforms (Takbiri et al., 2019), etc. Recent researches revealed that using gamification techniques and users' emotions in building interactive systems and recommender engines can significantly enhance system performance and increase user engagement (Feil et al., 2016; Rohani & Bastanfard, 2016). Gamification is the use of game components and rules like mechanics, dynamics, and emotions to engage audiences in non-game environments (Hamid & Kuppusamy, 2017; Robson et al., 2015). Increasing the accuracy of recommendations need more detail about users' tendencies and their needs.

Based on what has been explained above, gamification encourages users to be more interactive (Hassan & Hamari, 2020; Nguyen et al., 2018); these activities provide significantly important information for recommendation systems. This paper's key contribution is summarized as follows:

- Providing a thorough review of how gamified recommender systems may improve user engagement in various domains.
- In-depth research of gamified recommender systems, covering approaches, limits, evaluation metrics, results, datasets, domain areas, and recommendation methods.
- Providing a complete picture of the popularity of the subject, as well as any gaps and undiscovered areas.
- Introducing outstanding problems and potential future directions for researchers who are interested in the topic.

The rest of this paper is arranged in the following sections. Section 2 reviews previous related research and their difference from the current study. Discusses different techniques for building a recommender system as well as perusing their effect on user satisfaction and engagement. Moreover, it outlines fundamental concepts of gamification. Section 3 includes two main parts. The first describes the structure of gamified recommendation systems in detail as an intersection of gamification and recommenders. The second part comprises reviewing researches that specifically considered gamified recommendation systems as a proper answer for increasing user engagement. Section 4 presents in-depth surveying of the reviewed works from different aspects and detail such as approaches, limitations, evaluation metrics, proposed achievements, datasets, domain areas, and their recommendation techniques. Therefore, it introduces possible future research directions and open issues on this subject. Finally, Section 5 provides a summary of our findings and concludes our study.

## 2. Background

This section reviews related previous works that surveyed gamification strategies and recommenders. To the best of our knowledge, not a single study specifically focused on studying gamified recommendation systems to enrich user engagement in different domains. Therefore various methods for making a recommender system and their result on user satisfaction are investigated. Fundamental concepts of gamification, including different types and game elements, are also explained to fulfill the background knowledge about the topic.

### 2.1. Previous Work

In the past few years, some studies such as Klock et al. (Klock et al., 2020), Seaborn and Fels (Seaborn & Fels, 2015), Heilbrunn et al. (Heilbrunn et al., 2014), Khan et al. (Khan et al., 2021), Alhijawi and Yousef (Alhijawi & Kilani, 2020) and Tang et al. (Tang et al., 2013) were conducted to review and survey gamification or recommender systems. However, as a contribution, unlike the current work, no study specifically focused on reviewing the subject of boosting users' engagement with gamified recommenders in different environments and analyzing this topic's details. To provide a thorough study, we also reviewed mentioned researches and explained their notable aspects in the following lines.

Klock et al. (Klock et al., 2020) provided a comprehensive survey about the terminology of the game elements used in tailored gamification, dynamic modeling, and exploring multiple characteristics simultaneously. The main difference between the study and our research is providing an in-depth review by focusing on increasing users' engagement by employing gamification rules on recommendation systems in different statements. Seaborn and Fels (Seaborn & Fels, 2015) presented conceptual and practical findings from a systematic survey to explore gamification's theoretical and conceptual aspects and find distinct terms and concepts about the subject. The authors also tried to establish what links are between theoretical and applied works on gamification. In the gamification analytics field, Heilbrunn et al. (Heilbrunn et al., 2014) surveyed software solutions and assessed how well they met user demands. Khan et al. (Khan et al., 2021) conducted a thorough investigation on deep learning-based rating prediction and recommendation methods. The authors focused on rating prediction systems and looked at several techniques and designs. Alhijawi and Yousef (Alhijawi & Kilani, 2020) surveyed researches that used a recommender system in mobile, social network, or cloud environments. These recommender systems were categorized based on the technology or context in which they were used. Tang et al. (Tang et al., 2013) presented a review and classification of recommender systems by giving formal definitions of social recommendation. The authors discussed the property of social recommendation and its implications compared with those of traditional recommenders.

## 2.2. *Main Types of the Recommender Systems*

The main intention of a recommender system is to predict the user's tendency toward an item. In this way, many studies developed recommender systems with different functionalities to recommend books, movies, music, tourism destinations, friendships in social media, etc. (Jorro-Aragoneses et al., 2019; Roy & Ding, 2021). Lots of techniques have been used in various researches to build recommender systems (RS). However, the primary methods can be categorized into Content-Based, Context-Aware or Demographic-based, Knowledge-based, Collaborative Filtering, and Hybrid recommender systems. Among all of the five mentioned titles, collaborative filtering is the most common technique to build a recommender system (Esmaeili et al., 2020). An overview of the five mentioned criteria is shown in *Figure 2*.

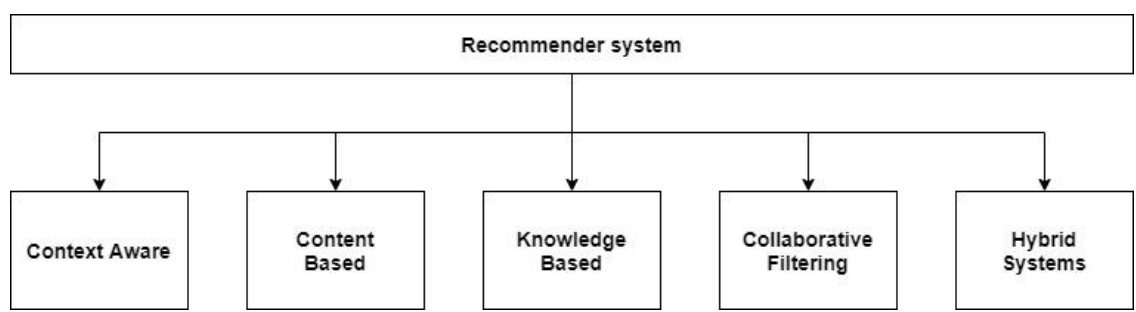

**Figure 2.** Recommender systems categorization based on their techniques.

Although each of the mentioned techniques is explained in detail in the following, to make this study more practical and beneficial, each subsection ended with discussing the advantages and drawbacks of that technique.

### 2.2.1. Context-aware

The primary data source in context-aware recommendation systems is contextual information such as the user's location, age, gender, occupation, nationality, etc (Raza & Ding, 2019). These recommenders usually categorize users into groups based on their similar properties and provide the same recommendations to each group's users. In general, context can be considered as any factor that was not explicitly defined; however, they affect the user's decision (Hosseinzadeh Aghdam, 2019; Villegas et al., 2018). The main elements of context-aware recommenders can be categorized as follows:

- **Time context**

In general, features that are related to the calendar can be considered as Time contexts. For example, days of a week, months, seasons, holidays, and even times of day can affect recommendations for different purposes. For instance, by considering the national holidays, a trip recommender system can change the popular items' weight to enhance recommendations' quality and users' satisfaction. Likewise, on hot days of the year in a specific area, an online shop's recommender engine can change the importance of relevant products such as sun creams by considering the changes of priority in the user's need.

- **Physical context**

Physical contexts are related to the user's surrounding area. An example of a context-aware recommender system is a music streaming website, which also recommends folkloric tracks by considering the user's location. In the mentioned example, while music tracks can be easily categorized by some related tags, more specialized techniques for analyzing voices and sounds in the environment can be used to improve the quality of the systems (Bastanfard et al., 2010).

- **Social context**

In the real world, social relations have a significant effect on the individual's behaviors (Liao & Yang, 2021). For example, by changing the marriage status of a user in a movie recommender system, the system can improve its users' satisfaction level by considering the target user as a member of the family cluster and recommend more relevant movies.

- **Modal context**

The model context is highly related to the user's mindset. A user's history is critical in this kind of context-aware recommender system. In an ideal step, the recommender can understand the user's purpose, for example, in online shopping, whether the user buys an item for personal use or as a gift. Having an excellent understating in modal can increase the quality of corporate communication.

*Advantages and Drawbacks of Context-Aware Recommenders*

Based on what has been explained above, context-aware recommender engines can provide personalized suggestions based on the user's background. This customization improves engagement and causes better communication. However, there are some critical concerns about context-aware recommenders. First, user's privacy can be in danger while such systems use data mining techniques to find more well-aimed features about users.

The second challenge ahead of context-aware recommender systems is calculation complexity. Considering more properties will increase the data dimension, and consequently, it causes more technical difficulties. Also, all the features and data can be changed dynamically over time. So the system has to check the precision of records continuously.

The third issue, which is also one of the main challenges for every kind of recommender system, is data sparsity (Choi et al., 2023). context-aware systems are highly related to the user's cooperation for sharing data, and lack of user activity causes some serious problems for this kind of recommender engine.

*2.2.2. Content-based*

In this technique, recommendations are obtained from what users chose in the past. For example, in a news content-based recommendation system, if a user likes an article about politics, the system will recommend other political news, that the user has not read yet. From a technical perspective, a content-based recommender engine uses a keyword-based vector-space structure with data from tags, comments, etc., to analyze the target user behaviors and then provide recommendations based on similarity matching scores with other contents (Pradhan

& Pal, 2020; Rahimpour Cami et al., 2019). Content-based recommender systems have a crucial role, especially in web 2.0 environments where suggestions are about posts, comments, and other multimedia material with tags and metadata (Keshtkar & Bastanfard, 2015; Lee & Lee, 2023). The architecture of the Content-based recommender systems in the basic form includes three modules (Lops et al., 2011). Each module is described below:

- **Content analyzer**

This module acts as a pre-processor and adopts the input data to a suitable format for the next processing component. The content can be an image, voice, textual document, etc (Bastanfard, Aghaahmadi, et al., 2009; Shang et al., 2021).

- **Profile learner**

In modern recommender systems, the Profile learner component highly benefits from machine learning techniques for collecting data and constructing the user profile. Likes and dislikes, ratings, and comments are some features that help the Profile learner component to increase the accuracy level of users' profiles (Jaffali et al., 2020).

- **Filtering component**

The final calculation about the target user's similarity and the dataset is done inside the Filtering component. The result of this component, which is also the final result of the system, is relevant recommendations (Izadi et al., 2010; Minoofam et al., 2022).

*Advantages and Drawbacks of Content-Based Recommenders*
The most advantageous feature of content-based recommender systems is the user's independence. Unlike the collaborative filtering systems that are dependent on other users' data for calculating the similarity score of the target user (Duma & Twala, 2018), in content-based recommenders, the system only needs the target user's information and content sources' metadata.

Second, in comparison with other types of recommender systems, content-based systems have more transparency and need simpler architecture and implementation. A noticeable issue in recommender systems is when a new user or a new item joined a system, and there is no history; this situation is known as the cold start problem. Cold start is more challenging, especially in collaborative filtering-based recommender systems, which need past interactions to calculate similarities between the entities. However, in content-based recommenders, cold start can happen only for users, as items can be recommended based on their tags (Lops et al., 2011; Narducci et al., 2016).

Although content-based systems have many advantages, there are some fundamental challenges with these kinds of systems. First, there should always be enough meta-data about each item. Otherwise, the Filtering component can not calculate similarity accurately (Mahtar et al., 2017). Another disadvantage is over-specialization, which means the system can lose its functionality in recommending different items and stick in a limited range of relevant items. This drawback is also known as the serendipity problem in recommender systems (Kotkov et al., 2016; Lops et al., 2011).

### 2.2.3. Knowledge-based

This kind of recommender system tries to improve the awareness of a knowledge base from different aspects of the system by extracting users' requirements and features. The well-trained knowledge base is then used to provide recommendations based on each user's characteristics by inference and reasoning techniques (Tejeda-Lorente et al., 2019). Knowledge-based recommender systems can be categorized into the two main below types (Aggarwal, 2016a):

- **Constraint-based recommender systems**

Some domain-specific rules are established in constraint-based systems, and the system makes suggestions based on them (Aggarwal, 2016b).

- **Case-based recommender systems**

Some similarity measures provide the domain knowledge utilized for giving weights to items in case-based systems (Bastanfard, Kelishami, et al., 2009).

*Advantages and Drawbacks of Knowledge-Based Recommenders*
Some notable advantages of knowledge-based recommenders are no dependency on large datasets, an excellent performance against adding new items, and cold start problems. Grey sheep attacks also are avoided (Bouraga et al., 2014; Khusro et al., 2016).
However, knowledge-based systems are not easy to implement. They need significant domain knowledge and expertise in knowledge representation.

### 2.2.4. Collaborative filtering

The collaborative filtering method is the most common of all the strategies for developing a recommender system described (Bobadilla et al., 2013; Verma & Aggarwal, 2020). The primary principle of collaborative filtering is to determine the similarity score between the target user and other users based on the user-item rating matrix by exploiting the relationship between users. The two primary kinds of collaborative filtering techniques are listed below (R et al., 2020):

- **Nearest neighbor collaborative filtering**

This approach chooses the most K-similar users to the target user and considers the average rating for them and is known as the k-nearest neighbors (KNN) algorithm (Soui & Haddad, 2023). If the neighborhood graph nodes are users, the method will be user-KNN, and if the nodes represent items, it will be item-KNN. In general, systems that work based on the neighborhood graph evaluate the values according to the nearby users or items.

- **Matrix factorization collaborative filtering**

Matrix factorization (MF) methods are latent factor models that consider user's past activities and create a history for each user (Rana & Jain, 2014). The system also uses the interaction of users about items for analyzing them. Singular Vector decomposition (SVD) is one of the most popular MF methods for building collaborative filtering recommender systems.

*Advantages and Drawbacks of Collaborative Filtering Recommenders*
Collaborative filtering is, as previously said, the most extensively utilized method for developing a recommender system. However, there are still two considerable drawbacks to these kinds of recommenders. The first drawback is about the time when a new user or a new item recently joined a system; this situation is known as the cold start problem (Adomavicius & Tuzhilin, 2005; Son, 2016). Another challenge ahead of collaborative filtering-based recommender systems is the data sparsity. Sparsity indicates the lack of users' interaction with the system. Users usually do not interact with systems. For example, do not participate in polls, writing comments, rating items, etc. (Ahmadian et al., 2019; Bobadilla et al., 2013).

### 2.2.5. Hybrid recommenders

A combination of two or more techniques for building a recommender system can provide a new hybrid method. The main goal of hybrid recommender systems is to increase the system's performance by profiting from the advantages of the basic techniques while overcoming their drawbacks. For instance, a popular method for building a hybrid recommender system is the fusion of content-based and collaborative filtering or context-aware (demographic) with collaborative filtering (demographic) (Riyahi & Sohrabi, 2020).

### 2.3. *Fundamnetal Concepts of Gamification*
Gamification offers several marketing benefits, and it has evolved into a multidisciplinary study field that is still in its infancy (Seaborn & Fels, 2015). Gamification may boost user engagement, leading to a higher conversion rate and, in turn, more income in the market (Hajarian et al., 2019; Takbiri et al., 2023). There are two sorts of gamification: meaningful and reward-based, dependent on the game features employed (Goshevski et al., 2017). Points, levels, leaderboards, and badges, generally known as game mechanics, are common game elements for reward-based gamification (Morschheuser et al., 2018). Challenges, incentives, societal influences, and user specifics will all be addressed using these game mechanisms . Play (exploration), choice, and information are all common game elements in meaningful gamification (Nicholson, 2015). Meaningful gamification boosts consumers' innate motivation, while reward-based gamification boosts their extrinsic motivation (Goshevski et al., 2017). Meaningful gamification may be combined with reward-based gamification to increase productivity and reach various goals. *Table 1* indicates the mentioned game elements followed by their explanations.

**Table. 1.** Game elements explanation

| Gamification type | Game elements | Explanation |
| --- | --- | --- |
| **Reward-based** | Points | Displays a user's score |
| | Levels | The degree of competence possessed by a user |
| | Leaderboards | Compares the status of a user to the status of other users |
| | Badges | As a reward, users will get an icon to display on their profile |
| **Meaningful** | Play | Providing a gamified environment in which a person can explore |
| | Choice | A user's capability of having different options from which to choose |
| | Information | Informing users about the consequences of their actions |

## 3. Gamified Recommender Systems

Many studies employed gamification techniques to implement personalized recommender systems with different purposes from turism to educational systems (Akhriza & Mumpuni, 2020; Nuanmeesri, 2022). Although assisting users through providing personalized recommendations has a significant impact on user satisfaction and engagement. However, while on one side, recommender engines need enough information about users to provide acceptable suggestions, on the other side, data sparsity is one of the fundamental challenges in recommender systems. As explained in the last sections, gamification methods are a superior solution to encourage users to have more interaction with the systems and provide more information. Based on this concept, a gamified recommender system as an intersection of gamification solutions and individualized recommendations may boost user engagement by providing a personalized gamified environment. To provide a clear view of the explained structure, the main possible gamified elements as an input to a gamified recommendation system with their relationships are visualized in *Figure 3*.

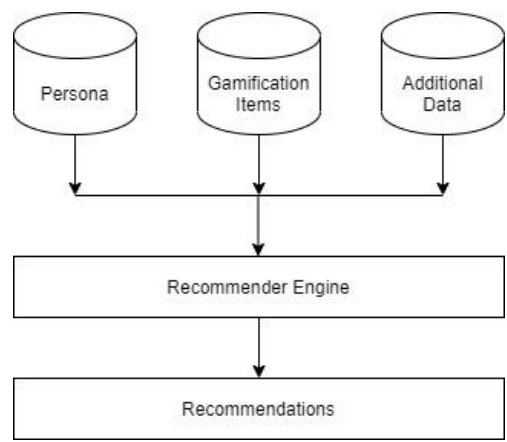

**Figure 3.** Fundamental structure of a gamified recommendation system with its input.

Persona, Gamification items, Additional Data, are the three main possible gamified elements for implementing a recommender system based on gamification rules (Tondello et al., 2017). The following paragraphs contain an explanation of each element with a real-world sample.

**Persona**
A common way to explore details about users is to direct questions through surveys and forms (Patrício et al., 2018). However, usually, users do not answer the question except the mandatory ones. In this situation, considering scores and titles for completed profiles can motivate users to reveal more details. *Figure 4* shows an example of asking basic information in Linkedin. Gender, age, education level, location are some of the typical info in creating personas. With over 690 million members from countries around the world, LinkedIn is one of the most extensive professional social networks. Job seekers may publish their CVs, and companies can advertise opportunities on the site (www.linkedin.com), which is used mainly for professional networking and career advancement (Redmond, 2022).

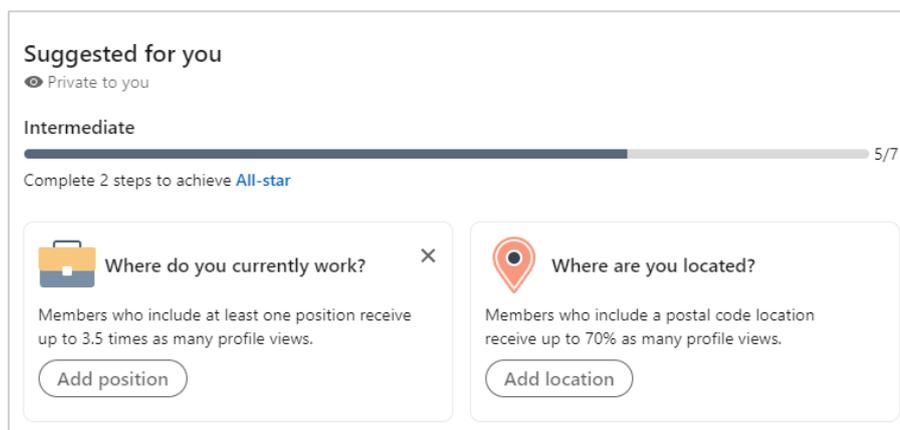

**Figure 4.** A gamification sample of using badges and levels and for motivating users to complete their profiles on Linkedin (*LinkedIn*).

Building an accurate persona of users can highly increase the quality of recommendations. As shown in *Figure 4*, LinkedIn users, by completing their profile detail step by step, assist the system's recommender engine in providing better suggestions based on their properties and needs.

**Gamification items**
A gamified system needs some items to run its gamification techniques. In a recommender system, when users have activities such as like, dislike, comment, buy, add, remove, etc the system can analyze the activity and provide some reactions (Al-Dhanhani et al., 2015). For instance, in an educational system, registering for a course can be the start of a gamified path that can be continued with achiving badges. *Figure 5* shows an example of badges on Kaggle. Kaggle is a data science and machine learning community on the internet. People may create virtual communities on the Kaggle site (www.kaggle.com), such as online notebooks and forums, to debate their models, feature choices, loss functions, and so on (Bojer & Meldgaard, 2021).

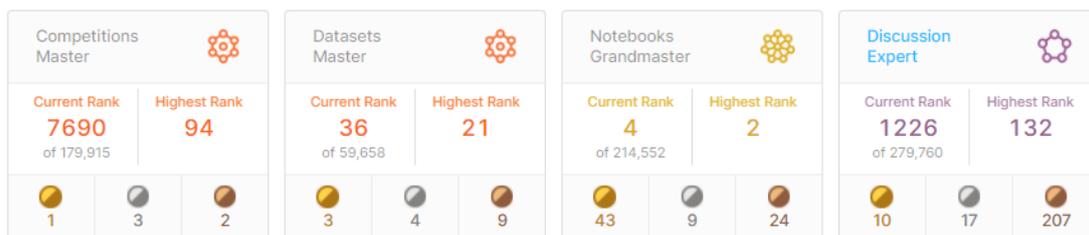

**Figure 5.** A gamification sample of using badges for users who were active in different section on Kaggle (*Kaggle*).

As *Figure 5* shows, for Kaggle users, the recommender engine can assist the target users in finding related items to their educational roadmap by finishing more courses and competitions.

**Additional Data**

Additional or contextual information can be considered as any information that users did not explicitly express (Mazarakis & Bräuer, 2023). For instance, in a tourism destination recommender, the system can ask users about their local time and location to customize the application's interface and switch between the day and night theme. Likewise, the recommender engine can suggest destinations near the user.

### 3.1. Related studies

The following paragraphs include reviewing the researches that specifically proposed the result of using gamified recommendation systems to improve users' engagement in different domains. An overview of the reviewed papers, including their approach and limitations, is summed up in the discussion section and *Table 2*.

Khoshkangini et al. (Khoshkangini et al., 2021) suggested a completely automated procedural content generation (PCG) and recommendation technique to provide motivating challenges for users. According to the findings, offering individualized, dynamic challenges based on the user's history, game state, and performance resulted in higher levels of engagement and completion.

Hajarian and Hemmati (Hajarian & Hemmati, 2020) presented a gamified word-of-mouth recommendation system that encourages users to promote things to one another, resulting in increased consumer sales and engagement. Users may ask questions regarding the items they have purchased using the suggested way. Responders get points by answering questions. By accumulating a particular number of points, users will be eligible to win rewards. There is also a leaderboard to provide a feeling of competitiveness. The findings point to an increase in client sales and unique visits.

To ensure a long-term involvement, Zhao et al. (Zhao et al., 2020) studied how to create customized and gamified recommender systems in health and fitness. Dynamic player modeling and wearable-based monitoring were introduced to provide customized game features and activity recommendations. The findings revealed that individualized recommendations based on player modeling and gamification might increase participants' motivation and participation in fitness activities.

González-González et al. (González-González et al., 2019) looked at how serious games, namely exergames, may be used in rehabilitation to inspire, engage, and promote patient adherence to therapy. They offered an intelligent exergame-based rehabilitation system, which comprises a platform with an exergame player and a designer tool. The proposed recommender system analyzes user interactions and history to suggest new gamified exercises based on different difficulty levels and user skills. KINECT sensors were used in this research to recognize motions without the need for physical contact. In the study, the effectiveness, efficiency, learnability, and satisfaction of gesture-based exercises were shown to be satisfactory and superior to traditional techniques. Furthermore, the recommender's behavior closely corresponds to what many experts may have determined when making a human judgment.

Cheng Kin et al. (Cheng Kin et al., 2018) looked at waste management as a possible solution to Malaysia's natural resource preservation concern. They emphasized the use of a framework to engage users with visual recommender systems and gamification to improve awareness in their daily household and industrial waste management practices. Gamification may drive people to learn, according to the research findings. As a result,

Octalysis (Chou, 2015) was developed as an appropriate gamification framework for human-centered design, and it was designed to optimize feelings, incentives, and task engagement.

Tondello et al. (Tondello et al., 2017) presented a framework for describing, designing, and assembling gamified recommender system components. Gameful systems may automatically offer tailored activities for each user to boost user motivation and engagement. This paper discusses items including game components and persuasive methods, users, transactions, and contextual information as possible inputs for a gamified recommender. The system's output also includes predicted ratings for each gameful action per user.

Su (Su, 2017) considered the learning environment and developed an adaptive learning route that included the Fuzzy Delphi Method, the Fuzzy ISM, and Kelly Repertory Grid Technology. They created a prototype gamification geometry-learning material module and an adaptable geometry-learning route diagram based on learning styles. The results suggest that the gamification geometry recommendation system, due to features like easy mobility, simple operation, and gamification entertainment surpasses typical learning course-guided recommendation mechanisms in satisfying users and inspiring students to study and practice the geometry unit on object shape, volume, circumference, and volume. As a result, assessment criteria like recall, precision, F1 index, and MAE have improved.

Khoshkangini et al. (Khoshkangini et al., 2017) presented a method for suggesting individualized and contextualized challenges that are customized to each player's preferences, history, game state, and performance based on Procedural Content Generation. The research was conducted in the field of smart cities, and the prototype was installed in Rovereto, Italy. The findings suggest that introducing customized playable material into a gamified system boosts long-term engagement and retention, hence magnifying gamification's persuasive power.

Herpich et al. (Herpich et al., 2017) looked at how gamification may be used to improve user acceptance of a context-aware recommender system for the elderly. The approach involves improving physical and psychological well-being by boosting the comfort and ease of one's living environment or by engaging in physical or mental activities. The findings show that users collaborated and were more satisfied with the suggested recommender system's activities.

Mulholland et al (Mulholland et al., 2015) employed sentiment analysis and gamification strategies in an education domain for recommending related videos. The research aimed to analyse the effect of sentiment analysis, emotions and gamified items on enhancing user satisfaction and engagement.

In the e-health domain, the paper proposed by Di Bitonto et al. (Di Bitonto et al., 2015) discusses several innovative learning environment options. The fundamental aim is to create engaging, individualized learning experiences based on the particular requirements of individual actors using a mix of social learning and game-based learning pedagogical techniques as well as social network and recommender system technology methods. Defining such learning models based on the provided outcomes helps boost user engagement and motivation, which are critical variables for effective learning.

Ziesemer et al. (CA Ziesemer et al., 2014) proposed a theoretical foundation for how gamification might promote ratings and enhance user involvement in activities like product rating on e-commerce platforms and help users overcome cold-start issues. The findings demonstrate that gamification may help improve recommendations, explicit feedback, goods, and the reputation of users.

Meder et al. (Meder et al., 2013) looked at how workers feel about implementing a gamified system at workplaces. First, an online poll was used to assess consumers' attitudes about gamification in the workplace. The users' subjective gamification experience was then compared to their actual behavior using a gamified Social Enterprise Bookmarking System. After that, the participants' interaction with the system was studied. The findings show a significant link between users' acknowledgment of gamification and their actual interactions with one.

López-Rodríguez and García-Linares (López-Rodríguez & García-Linares, 2013) offered a technology approach for cognitive rehabilitation that was motor-based, remote, and included learning and gamification. A patient's score in a game is compared to the scores of other patients who are undergoing the same sort of rehabilitation. To maximize the effectiveness of the rehabilitation plan, a recommendation system will be able to customize it to each patient's requirements and progression. Using artificial intelligence and data mining methods, researchers will be able to analyze data from a large number of patients, uncover trends, and construct models that display the ideal games and itineraries for each patient, based on their problems, age, gender, and other factors. Because of innovations like customized suggestions and social feedback systems in the gamification platform, the system will be more successful rehabilitation treatments than usual alternatives, according to the supplied arguments.

Silva et al. (Silva et al., 2013) established a rating system for the environment based on their evaluation of its long-term viability. The ranking of people and environments is then determined using similarity and clustering techniques. This study investigates the effects of gamification and knowledge dissemination in communities of intelligent settings on the behavioral and physical transformation. The transmission of this information allows users within smart environment communities to make use of it and improve the overall operation of the community. Because of what has been explained, gamification may increase user loyalty and keep them engaged in the goal by making it more entertaining.

## 4. Discussion

This section includes the study's results and expressly presents a comprehensive analysis of using gamified recommender systems to increase engagement and satisfaction by analyzing the details from some of the essential views that can be applied to the topic. A condensed comparison based on the surveyed approach and their limitations is presented in *Table 2*. Based on the limitation column values, it can be concluded that evaluation and lack of sufficient experimental results is a considerable limitation in the reviewed papers.

**Table. 2.** An overview of the reviewed papers and their approach and limitations.

| Name | Approach | Limitation |
|---|---|---|
| (Nuanmeesri, 2022) | Considering users' travel history and encouraging them by gamification techniques to share more data about visited locations. Entrepreneurs, based on this data, can recommend destinations to travellers. | • Evaluation only based on a customized dataset<br>• No comparision with other state of the art approaches |
| (Khoshkangini et al., 2021) | Automatically generate individualized and contextualized tasks and challenges for each player based on their preferences, history, game state, and performances. | Evaluation period (3-week experiment) |
| (Hajarian & Hemmati, 2020) | Proposing a recommender enhancement solution that incorporates gamification in electronic word of mouth. | • Evaluation period (32 days)<br>• Number of participants (84 member) |
| (Zhao et al., 2020) | Created a new player model and associated system architecture for a gamified fitness activity recommender system. | Number of participants (40 participants, including 23 men and 17 women.) |
| (González-González et al., 2019) | Based on the characteristics of each user or group of users in rehabilitation, personalized exercises are designed and advised. To identify gestures, the system uses KINECT sensors. | • Number of participants<br>• Lack of common evaluation metrics for recommender systems such as MAE, RMSE<br>• Subjective evaluation based on users' and experts' feedbacks |
| (Cheng Kin et al., 2018) | Using a gamification framework to engage users with a visual recommender system and gamification to raise waste management awareness. | No actual implementation and evaluation, only conceptual |
| (Tondello et al., 2017) | Proposed a generic architecture for creating gamification-related recommender system components. | No actual implementation and evaluation, only theoretical |
| (Su, 2017) | Proposing a gamified geometry teaching material based on a hybrid adaptive learning recommender. | Only the geometry unit in G5 and G6 mathematics instruction is used. |
| (Khoshkangini et al., 2017) | Using a gamification framework, creating and suggesting customized challenges for the Smart Cities domain. | • Evaluation period (9 weeks)<br>• Number of participants (110 active players)<br>• Number of feedbacks in an in–App survey (36 users)<br>• Location limitations (Only on Rovereto, Italy) |
| (Herpich et al., 2017) | Using gamification to improve user acceptance of the context-aware recommender system. | • Gender unbalance<br>    Group1(11 women, 1man)<br>    Group2(8 women, 1man) |
| (Mulholland et al., 2015) | Employing sentiment analysis and gamification strategies to recommend educational videos. | • Number of evaluation data (39 sentences) |
| (Di Bitonto et al., 2015) | Incorporating social learning and game-based educational methodologies with technological tools from social networks and recommendation systems. | • Number of attendees (Departments of Nephrology and Cardiology "Policlinico" hospital in Bari: 16 doctors, six nurses and 20 patients, and carers at home)<br>• Require a longer-term study |
| (CA Ziesemer et al., 2014) | offering an online poll to determine the benefits of gamification on enhancing recommendations and overcoming cold-start problems. | • No implementation of the framework |
| (Meder et al., 2013) | Looking at the function of gamification in the workplace from employees' perspective, first via a questionnaire and then through a social business bookmarking system. | • Number of participants (84 members) |
| (López-Rodríguez & García-Linares, 2013) | Proposing a technology solution based on gamification and a recommender system for Cognitive Rehabilitation Motor, Remote. | • No Implementation |
| (Silva et al., 2013) | Gamification components encourage users to rate surroundings based on their sustainability evaluation and provide ideas for improvement. | • Number of tests<br>• Lack of details about using gamification |

**Table. 3.** An overview of the reviewed papers based on their evaluation metrics, proposed achievements and dataset.

| Name | Evaluation Metrics | Proposed Achievements | Dataset |
|---|---|---|---|
| (Nuanmeesri, 2022) | • Black box testing by experts<br>• t-Test<br>• Users' feedbacks | • Better sensitivity<br>• Higher accuracy | Customized for this research |
| (Khoshkangini et al., 2021) | • Acceptance rate of challenges<br>• Improvement in mobility habits<br>• Economicity | • Better completion rate<br>• Higher level of improvement<br>• Improvement for the same amount of per capita reward | Customized for this research |
| (Hajarian & Hemmati, 2020) | • Website visits<br>• Number of purchases | • Increasing the customers' purchases and unique visits<br>• The number of female consumers who visited climbed by 116%, while the number of male customers who visited increased by 100%.<br>• The count of purchases by customers has increased. | A cosmetics e-commerce website |
| (Zhao et al., 2020) | Period of 60 days<br>• Motivation<br>• Satisfaction<br>• Preference | • Increased motivation<br>• Increased satisfaction<br>• Increased preference | Customized for this research |
| (González-González et al., 2019) | For gesture-based exercises:<br>• Effectiveness<br>• Efficacy<br>• Learnability<br>• Satisfaction | • Better than traditional exercises in effectiveness, efficiency, learnability, and satisfaction<br>• Suggestion was closely mirroring the experts' decisions | Customized for this research |
| (Cheng Kin et al., 2018) | • No Evaluation (Only theoretical) | • Expect after implementation can be helpful in the intent domain. | No actual implementation and evaluation, only conceptual |
| (Tondello et al., 2017) | • No Evaluation (Only theoretical) | • Proposing the framework as the first discussion of how to use recommender systems in personalized gamification | No actual implementation and evaluation, only theoretical |
| (Su, 2017) | • Questioner<br>• Recall<br>• Precision<br>• F1 index<br>• MAE | • Better outcomes<br>• Satisfaction scores higher than 90%<br>• Recall (95%)<br>• Precision (68%)<br>• F1 index (45%)<br>• MAE (8%) | Customized for this research |
| (Khoshkangini et al., 2017) | • User's participating effects on the city<br>• In-App survey<br>• Precision<br>• Recall<br>• Accuracy | • Challenge acceptance ratio of 24%<br>• Fulfillment of at least 50% of the challenge goal<br>• Precision: 31%<br>• Recall: 64%<br>• Accuracy: 37% | Customized for this research |
| (Herpich et al., 2017) | • Comparing user's activities with and without recommendations | • Increased user appreciation | Customized for this research |
| (Mulholland et al., 2015) | • Customized | • May cause satisfaction | Customized for this research |
| (Di Bitonto et al., 2015) | • Users' feedbacks | • Practitioners give it high marks in terms of information gained and user satisfaction. | Customized for this research |
| (CA Ziesemer et al., 2014) | • An online survey, 367 participants (221 male, 146 female) | • Presents a theoretical solution for the cold-start problem | Customized for this research |
| (Meder et al., 2013) | • Comparing user's activities with the proposed systems and answers in the questionnaire | • A strong relationship between employees' perception of gamification and their actual interaction | Customized for this research |
| (López-Rodríguez & García-Linares, 2013) | • Providing a social-powered gamification platform for motor rehabilitation to improve in-house rehabilitation | • The study is a project proposal | No dataset yet |
| (Silva et al., 2013) | • Comparing sustainability recommendations in simulated environments | • Using recommendations on the living room alone was sufficient to improve the target environment sustainability index. | Customized for this research |

*Table 3* provides an overview of the reviewed papers based on their evaluation metrics, proposed achievements, and dataset. From the results, it is observed that except for the standard classification accuracy metrics such as Recall and Precision that were used in some of the mentioned studies, most of the researchers tried to evaluate their approach and methods by analyzing the user's activities and feedback. Moreover, in contrast to the other recommender systems application domains, which have some baseline datasets for running experimental results, there is no such typical dataset in the area of gamified recommenders for improving engagement.

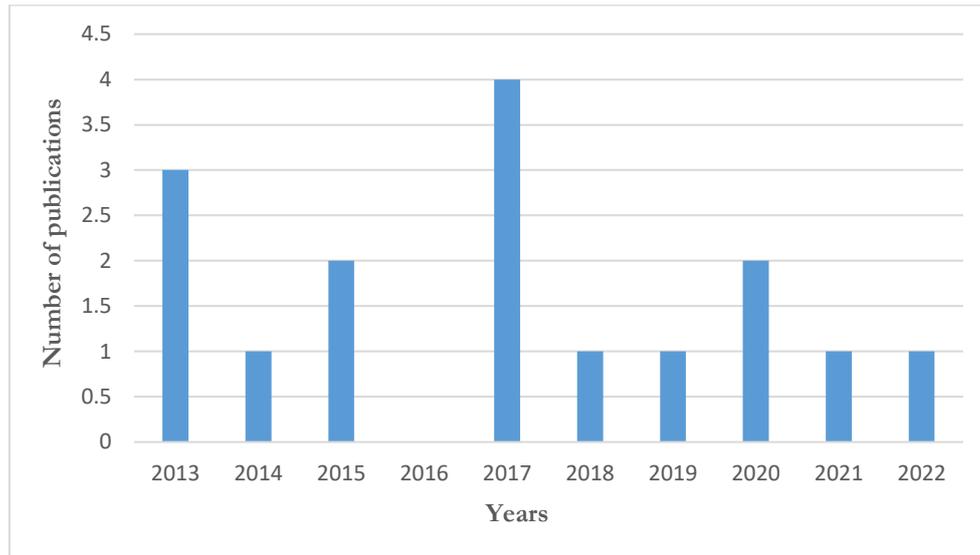

**Figure 6.** Distribution of reviewed papers based on their publishing date.

As shown in *Figure 6*, Years 2017 and 2013 have the most significant number of published papers about enhancing user engagement by employing gamified recommender systems. Respectively, 2015 and 2020 are the next ones.

**Table. 4.** Reviewed papers and their Application Domain

| Name | Application Domain |
|---|---|
| (Nuanmeesri, 2022) | Healthcare and Fitness |
| (Khoshkangini et al., 2021) | Healthcare and Fitness |
| (Hajarian & Hemmati, 2020) | E-commerce |
| (Zhao et al., 2020) | Healthcare and Fitness |
| (González-González et al., 2019) | Healthcare and Fitness |
| (Cheng Kin et al., 2018) | Smart City |
| (Tondello et al., 2017) | Educational |
| (Su, 2017) | E-Learning |
| (Khoshkangini et al., 2017) | Healthcare and Fitness |
| (Herpich et al., 2017) | Healthcare and Fitness |
| (Mulholland et al., 2015) | Educational |
| (Di Bitonto et al., 2015) | Healthcare and Fitness |
| (CA Ziesemer et al., 2014) | E-commerce |
| (Meder et al., 2013) | Educational |
| (López-Rodríguez & García-Linares, 2013) | Healthcare and Fitness |
| (Silva et al., 2013) | Smart City |

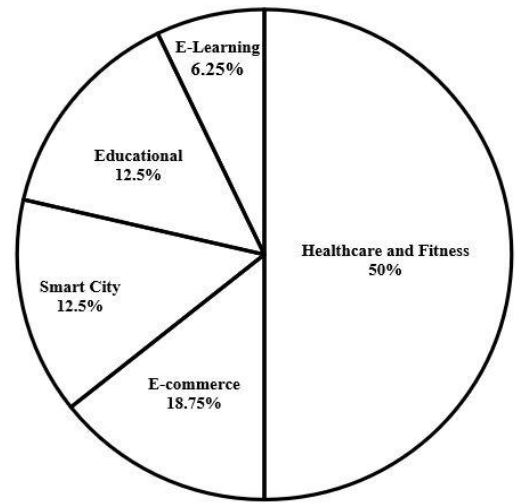

**Figure 7.** Distribution of the application domains in the reviewed papers.

*Figure 7* shows a pie chart that presents another view of the given values in *Table 4* about the distribution of the application domains in the reviewed papers. As a consequence of these values, it can be concluded that Healthcare and Fitness is the most common application domain in gamified recommenders to improve users' satisfaction and engage them in the defined scenarios.

Another categorization in this work was done based on the method used to implement the recommender system in the reviewed papers. *Table. 5* shows the distribution of these studies based on their recommendation technique. It is clear that hybrid systems based on collaborative filtering and context-aware techniques are the most common approach in building recommenders in the domain: gamification and user engagement.

**Table. 5.** Reviewed papers categorization based on their recommendeation system technique

| Name | Recommender system type |
|---|---|
| (Nuanmeesri, 2022) | Hybrid(Content-Based and Context Aware) |
| (Khoshkangini et al., 2021) | Hybrid(Content-Based and Context Aware) |
| (Hajarian & Hemmati, 2020) | Collaborative Filtering |
| (Zhao et al., 2020) | Hybrid(Content-Based and Context Aware) |
| (González-González et al., 2019) | Hybrid(Collaborative Filtering and Context Aware) |
| (Cheng Kin et al., 2018) | Content-Based |
| (Tondello et al., 2017) | General Framework |
| (Su, 2017) | Hybrid(Collaborative Filtering and Content-Based) |
| (Khoshkangini et al., 2017) | Hybrid(Collaborative Filtering and Context Aware) |
| (Herpich et al., 2017) | Hybrid(Collaborative Filtering and Context Aware) |
| (Mulholland et al., 2015) | Hybrid(Collaborative Filtering and Knowledge-Based) |
| (Di Bitonto et al., 2015) | Context Aware |
| (CA Ziesemer et al., 2014) | Hybrid(Collaborative Filtering and Content-Based) |
| (Meder et al., 2013) | Hybrid(Collaborative Filtering and Content-Based) |
| (López-Rodríguez & García-Linares, 2013) | Hybrid(Collaborative Filtering and Context Aware) |
| (Silva et al., 2013) | Collaborative Filtering |

*4.1. Open issues and possible directions*

Based on the in-depth analyses discussion about the various aspects of the topic, it can be clearly concluded that the area of applying gamified recommender systems with the goal of increasing user engagement is wide open for further investigation. The following are some examples of possible study directions in this field.

- As a consequence of the findings, the majority of researchers attempted to assess their strategy and techniques by monitoring the user's behaviors and feedback. In further research, conventional classification accuracy measures such as Recall and Precision can be used to provide a more clear view of the results.
- In contrast to the other recommender systems application areas, which usually have some baseline datasets for running experimental findings, there is no such typical dataset in the field of gamified recommenders for enhancing engagement, which can be considered as a significant limitation. As a result, having a robust synthetic dataset with easy and fast access seems essential for performing advanced analytics on gamified approaches.
- Employing a gamified recommender in other domains such as requirement engineering to assist experts and stakeholders on requirement elicitation activities would also be benefical.
- Therefore, using unsupervised machine learning techniques to cluster users based on their features and provide more innovative gamified recommendation system can be another direction for our future work.

## 5. Conclusion

The digital age has affected many perspectives of industries. New channels of communication with users are one of the most notable changes in the modern era. Today, companies can effectively improve their brand's popularity in social networks and increase user engagement. Selling products continuously without any time or geographical limitations by a website caused fundamental annual business profit changes.

Among many ways to increase user engagement and effective communication, recommender systems by providing highly personalized suggestions can play crucial roles. There are lots of approaches for building an efficient recommender system such as content-based, context-aware, knowledge-based, collaborative filtering, and hybrid methods. In order to boost the accuracy of a recommender engine, the system needs detailed information from its users. Based on this concept using gamification techniques can highly increase user's interaction and engagement with the system.

In this study, first, previous related research and their difference from the current study were reviewed. Next, different techniques for building a recommender system were explained by perusing their effect on user satisfaction and engagement. Then, fundamental concepts of gamification were proposed. After that, the gamified recommender is explained as an intersection of the gamification and recommendation system. Moreover, researches that are specifically considered gamified recommendation systems as a proper answer for increasing engagement were surveyed. Finally, reviewed works were comprehensively analyzed from diverse factors such as approaches, limitations, evaluation metrics, proposed achievements, datasets, domain areas, and their recommendation techniques. Therefore, open issues and possible directions were described for researchers interested in this subject.

# References

Abdollahpouri, H., Adomavicius, G., Burke, R., Guy, I., Jannach, D., Kamishima, T., Krasnodebski, J., & Pizzato, L. (2020). Multistakeholder recommendation: Survey and research directions. *User Modeling and User-Adapted Interaction* https://doi.org/10.1007/s11257-019-09256-1

Adomavicius, G., & Tuzhilin, A. (2005). *Toward the next generation of recommender systems: a survey of the state-of-the-art and possible extensions* IEEE Transactions on Knowledge and Data Engineering,

Aggarwal, C. C. (2016a). Knowledge-Based Recommender Systems. In *Recommender Systems* (pp. 167-197). Springer, Cham. https://doi.org/10.1007/978-3-319-29659-3_5

Aggarwal, C. C. (2016b). *Recommender systems* (Vol. 1). Springer.

Ahmadian, S., Afsharchi, M., & Meghdadi, M. (2019). A novel approach based on multi-view reliability measures to alleviate data sparsity in recommender systems. *Multimedia Tools and Applications* 17763–17798. https://doi.org/10.1007/s11042-018-7079-x

Akhriza, T. M., & Mumpuni, I. D. (2020). Gamification of the Lecturer Career Promotion System with a Recommender System. 2020 Fifth International Conference on Informatics and Computing (ICIC),

Al-Dhanhani, A., Mizouni, R., Otrok, H., & Al-Rubaie, A. (2015). Analysis of collaborative learning in social network sites used in education. *Social Network Analysis and Mining*, *5*(1), 1-18.

Alhijawi, B., & Kilani, Y. (2020). The recommender system: a survey. *International Journal of Advanced Intelligence Paradigms*, *15*(3), 229-251.

Bastanfard, A., Aghaahmadi, M., Fazel, M., & Moghadam, M. (2009). Persian viseme classification for developing visual speech training application. Pacific-Rim Conference on Multimedia,

Bastanfard, A., Kelishami, A. A., Fazel, M., & Aghaahmadi, M. (2009). A comprehensive audio-visual corpus for teaching sound persian phoneme articulation. 2009 IEEE International Conference on Systems, Man and Cybernetics,

Bastanfard, A., Rezaei, N. A., Mottaghizadeh, M., & Fazel, M. (2010). A novel multimedia educational speech therapy system for hearing impaired children. Pacific-Rim Conference on Multimedia,

Bobadilla, J., Ortega, F., Hernando, A., & Gutiérrez, A. (2013). Recommender systems survey. *Knowledge-Based Systems*, *46*, 109-132.

Bojer, C. S., & Meldgaard, J. P. (2021). Kaggle forecasting competitions: An overlooked learning opportunity. *International Journal of Forecasting*, *37*(2), 587-603.

Bouraga, S., Jureta, I., Faulkner, S., & Herssensm, C. (2014). Knowledge-based recommendation systems:A survey. *International Journal of Intelligent Information Technologies*, *10*(2), 1-19. https://doi.org/10.4018/ijiit.2014040101

CA Ziesemer, A. d., Müller, L., & Silveira, M. S. (2014). Just rate it! Gamification as part of recommendation. International Conference on Human-Computer Interaction,

Camilleri, M. (2015). Valuing Stakeholder Engagement and Sustainability Reporting. *Corporate Reputation Review* https://doi.org/10.1057/crr.2015.9

Cheng Kin, M., Junita Shariza, M. N., Ah Choo, K., & Ulka Chandini, P. (2018). *Waste Management Mobile Application via Virtualisation recommender and Gamification framework* Virtual Informatics International Seminar 2018 (VIIS2018), Bangi, Malaysia.

Choi, S.-M., Lee, D., Jang, K., Park, C., & Lee, S. (2023). Improving Data Sparsity in Recommender Systems Using Matrix Regeneration with Item Features. *Mathematics*, *11*(2), 292.

Chou, d. Y.-k. (2015). *Actionable Gamification-Beyond Points, Badges, and Leaderboards. Octalysis Media*. CreateSpace Independent Publishing Platform, USA.

Conaway, R., & Garay, M. C. (2014). Gamification and service marketing. *SpringerPlus*, *3*(1), 1-11.

Davidaviciene, V., Pabedinskaite, A., & Davidavicius, S. (2017). Social networks in B2B and b2c communication. *International Journal of Scholarly Papers - Transformations in Business & Economics*, *16*.

Di Bitonto, P., Pesare, E., Rossano, V., & Roselli, T. (2015). Smart learning environments using social network, gamification and recommender system approaches in e-health contexts. In *Smart education and smart e-learning* (pp. 491-500). Springer.

Duma, M., & Twala, B. (2018). Optimising latent features using artificial immune system in collaborative filtering for recommender systems. *Applied Soft Computing*, *71*, 183-198.
Esmaeili, L., Mardani, S., Hashemi Golpayegani, S. A., & Zanganeh Madar, Z. (2020). A novel tourism recommender system in the context of social commerce. *Expert Systems with Applications*, *149*.
Feil, S., Kretzer, M., Werder, K., & Maedche, A. (2016). Using gamification to tackle the cold-start problem in recommender systems. Proceedings of the 19th ACM Conference on Computer Supported Cooperative Work and Social Computing Companion,
Fombrun, C. J. (1996). Reputation : realizing value from the corporate image. *Boston Mass by Harvard Business School Press*.
González-González, C. S., Toledo-Delgado, P. A., Muñoz-Cruz, V., & Torres-Carrion, P. V. (2019). Serious games for rehabilitation: Gestural interaction in personalized gamified exercises through a recommender system. *Journal of biomedical informatics*, *97*, 103266.
Goshevski, D., Veljanoska, J., & Hatziapostolou, T. (2017). A review of gamification platforms for higher education. Proceedings of the 8th Balkan Conference in Informatics,
Hajarian, M., Bastanfard, A., Mohammadzadeh, J., & Khalilian, M. (2019). A personalized gamification method for increasing user engagement in social networks. *Social Network Analysis and Mining*, *9*(1), 1-14.
Hajarian, M., & Hemmati, S. (2020). A gamified word of mouth recommendation system for increasing customer purchase. 2020 4th International Conference on Smart City, Internet of Things and Applications (SCIOT),
Hamid, M., & Kuppusamy, M. (2017). Gamification implementation in service marketing: a literature. *Electronic Journal of Business & Management*, *2*(1), 38-50.
Hassan, L., & Hamari, J. (2020). Gameful civic engagement: A review of the literature on gamification of e-participation. *Government Information Quarterly*, *37*(3), 101461.
Heilbrunn, B., Herzig, P., & Schill, A. (2014). Tools for gamification analytics: A survey. 2014 IEEE/ACM 7th international conference on utility and cloud computing,
Herpich, M., Rist, T., Seiderer, A., & André, E. (2017). Towards a gamified recommender system for the elderly. Proceedings of the 2017 International Conference on Digital Health,
Hosseinzadeh Aghdam, M. (2019). Context-aware recommender systems using hierarchical hidden Markov model. *Physica A: Statistical Mechanics and its Applications*, *518*, 89-98. https://doi.org/10.1016/j.physa.2018.11.037
Izadi, A., Rezaei, M., & Bastanfard, A. (2010). A computerized method to generate complex symmetric and geometric tiling patterns. In *Intelligent Computer Graphics 2010* (pp. 185-210). Springer.
Jaffali, S., Jamoussi, S., Smaïli, K., & Ben Hamadou, A. (2020). Like-tasted user groups to predict ratings in recommender systems. *Social Network Analysis and Mining*, *10*(1), 1-11.
Jorro-Aragoneses, J. L., Recio-García, J. A., Díaz-Agudo, B., & Jimenez-Díaz, G. (2019). RecoLibry-core: A component-based framework for building recommender systems. *Knowledge-Based Systems*, *182*.
*Kaggle*.  Retrieved 1 April from https://www.kaggle.com/
Kartajaya, H., Kotler, P., & Hooi, D. H. (2019). Marketing 4.0: moving from traditional to digital. *World Scientific Book Chapters*, 99-123.
Keshtkar, M., & Bastanfard, A. (2015). Determining the best proportion of music genre to be played in a radio program. 2015 7th Conference on Information and Knowledge Technology (IKT),
Khan, Z. Y., Niu, Z., Sandiwarno, S., & Prince, R. (2021). Deep learning techniques for rating prediction: a survey of the state-of-the-art. *Artificial Intelligence Review*, *54*(1), 95-135.
Khoshkangini, R., Valetto, G., & Marconi, A. (2017). Generating personalized challenges to enhance the persuasive power of gamification. Personalization in Persuasive Technology Workshop,
Khoshkangini, R., Valetto, G., Marconi, A., & Pistore, M. (2021). Automatic generation and recommendation of personalized challenges for gamification. *User Modeling and User-Adapted Interaction*, *31*(1), 1-34.

Khusro, S., Ali, Z., & Ullah, I. (2016). Recommender Systems: Issues, Challenges, and Research Opportunities. *Information Science and Applications (ICISA) 2016*, 1179-1189. https://doi.org/10.1007/978-981-10-0557-2_112

Klock, A. C. T., Gasparini, I., Pimenta, M. S., & Hamari, J. (2020). Tailored gamification: A review of literature. *International Journal of human-computer studies*, *144*, 102495.

Kotkov, D., Wang, S., & Veijalainen, J. (2016). A survey of serendipity in recommender systems. *Knowledge-Based Systems*, *111*. https://doi.org/10.1016/j.knosys.2016.08.014

Lee, S. Y., & Lee, S. W. (2023). Normative or effective? The role of news diversity and trust in news recommendation services. *International Journal of Human–Computer Interaction*, *39*(6), 1216-1229.

Liao, S.-H., & Yang, C.-A. (2021). Big data analytics of social network marketing and personalized recommendations. *Social Network Analysis and Mining*, *11*(1), 1-19.

*LinkedIn*. Retrieved 1 April from https://www.linkedin.com/

López-Rodríguez, D., & García-Linares, A. (2013). Spare: Spatial rehabilitation with learning, recommendations and gamification. *ICERI2013 Proc*.

Lops, P., de Gemmis, M., & Semeraro, G. (2011). Content-based Recommender Systems: State of the Art and Trends. In *Recommender Systems Handbook* (pp. 73-105). https://doi.org/10.1007/978-0-387-85820-3_3

Lucassen, G., & Jansen, S. (2014). Gamification in consumer marketing-future or fallacy? *Procedia-Social and Behavioral Sciences*, *148*, 194-202.

Mahtar, S., Masrom, S., Omar, N., Khairudin, N., Nor Abdul Rahim, S. K., & Rizman, Z. I. (2017). Trust aware recommender system with distrust in different views of trusted users. *Journal of Fundamental and Applied Sciences*, *9*.

Mazarakis, A., & Bräuer, P. (2023). Gamification is working, but which one exactly? Results from an experiment with four game design elements. *International Journal of Human–Computer Interaction*, *39*(3), 612-627.

Meder, M., Plumbaum, T., & Hopfgartner, F. (2013). Perceived and actual role of gamification principles. 2013 IEEE/ACM 6th International Conference on Utility and Cloud Computing,

Minoofam, S. A. H., Bastanfard, A., & Keyvanpour, M. R. (2022). RALF: an adaptive reinforcement learning framework for teaching dyslexic students. *Multimedia Tools and Applications*, 1-24.

Morschheuser, B., Hassan, L., Werder, K., & Hamari, J. (2018). How to design gamification? A method for engineering gamified software. *Information and Software Technology*, *95*, 219-237.

Mulholland, E., Mc Kevitt, P., Lunney, T., Farren, J., & Wilson, J. (2015). 360-MAM-Affect: Sentiment analysis with the Google prediction API and EmoSenticNet. 2015 7th international conference on intelligent technologies for interactive entertainment (INTETAIN),

Narducci, F., Basile, P., Musto, C., Lops, P., Caputo, A., de Gemmis, M., Iaquinta, L., & Semeraro, G. (2016). Concept-based item representations for a cross-lingual content-based recommendation process. *Information Sciences*, *374*, 15-31. https://doi.org/10.1016/j.ins.2016.09.022

Nguyen, T. S., Mohamed, S., & Panuwatwanich, K. (2018). Stakeholder Management in Complex Project: Review of Contemporary Literature. *Journal of Engineering, Project, and Production Management*, 75-89.

Nicholson, S. (2015). A recipe for meaningful gamification. In *Gamification in education and business* (pp. 1-20). Springer.

Nuanmeesri, S. (2022). Development of community tourism enhancement in emerging cities using gamification and adaptive tourism recommendation. *Journal of King Saud University-Computer and Information Sciences*, *34*(10), 8549-8563.

Patrício, R., Moreira, A. C., & Zurlo, F. (2018). Gamification approaches to the early stage of innovation. *Creativity and Innovation Management*, *27*(4), 499-511.

Pradhan, T., & Pal, S. (2020). CNAVER: A Content and Network-based Academic VEnue Recommender system. *Knowledge-Based Systems*, *189*. https://doi.org/10.1016/j.knosys.2019.105092

R, K., Kumar, P., & Bhasker, B. (2020). DNNRec: A novel deep learning based hybrid recommender system. *Expert Systems with Applications*, *144*. https://doi.org/10.1016/j.eswa.2019.113054

Rahimpour Cami, B., Hassanpour, H., & Mashayekhi, H. (2019). User preferences modeling using dirichlet process mixture model for a content-based recommender system. *Knowledge-Based Systems*, 644-655.

Rana, C., & Jain, S. K. (2014). An extended evolutionary clustering algorithm for an adaptive recommender system. *Social Network Analysis and Mining*, *4*(1), 1-13.

Rapp, A. (2020). An exploration of world of Warcraft for the gamification of virtual organizations. *Electronic Commerce Research and Applications*, *42*, 100985.

Raza, S., & Ding, C. (2019). Progress in context-aware recommender systems — An overview. *Computer Science Review*, *31*, 84-97. https://doi.org/10.1016/j.cosrev.2019.01.001

Redmond, W. (2022). *LinkedIn Business Highlights from Microsoft's FY22 Q2 Earnings*. https://news.linkedin.com/2020/april/linkedin-business-highlights-from-microsoft-s-fy20-q3-earnings

Riyahi, M., & Sohrabi, M. K. (2020). Providing effective recommendations in discussion groups using a new hybrid recommender system based on implicit ratings and semantic similarity. *Electronic Commerce Research and Applications*, *40*. https://doi.org/10.1016/j.elerap.2020.100938

Robson, K., Plangger, K., Kietzmann, J. H., McCarthy, I., & Pitt, L. (2015). Is it all a game? Understanding the principles of gamification. *Business horizons*, *58*(4), 411-420.

Rohani, A. R., & Bastanfard, A. (2016). Algorithm for persian text sentiment analysis in correspondences on an e-learning social website. *Journal of Research in Science, Engineering and Technology*, *4*(01), 11-15.

Roy, D., & Ding, C. (2021). Multi-source based movie recommendation with ratings and the side information. *Social Network Analysis and Mining*, *11*(1), 1-20.

Seaborn, K., & Fels, D. I. (2015). Gamification in theory and action: A survey. *International Journal of human-computer studies*, *74*, 14-31.

Shang, L., Zhang, D. Y., Shen, J., Marmion, E. L., & Wang, D. (2021). CCMR: A Classic-enriched Connotation-aware Music Retrieval System on Social Media with Visual Inputs. *Social Network Analysis and Mining*, *11*(1), 1-14.

Silva, F., Analide, C., Rosa, L., Felgueiras, G., & Pimenta, C. (2013). Social networks gamification for sustainability recommendation systems. In *Distributed Computing and Artificial Intelligence* (pp. 307-315). Springer.

Son, L. H. (2016). Dealing with the new user cold-start problem in recommender systems: A comparative review. *Information Systems*, *58*, 87-104. https://doi.org/10.1016/j.is.2014.10.001

Soui, M., & Haddad, Z. (2023). Deep learning-based model using DensNet201 for mobile user interface evaluation. *International Journal of Human–Computer Interaction*, *39*(9), 1981-1994.

Su, C. (2017). Designing and developing a novel hybrid adaptive learning path recommendation system (ALPRS) for gamification mathematics geometry course. *Eurasia Journal of Mathematics, Science and Technology Education*, *13*(6), 2275-2298.

Takbiri, Y., Amini, A., & Bastanfard, A. (2019). A Structured Gamification Approach for Improving Children's Performance in Online Learning Platforms. 2019 5th Iranian conference on signal processing and intelligent systems (ICSPIS),

Takbiri, Y., Bastanfard, A., & Amini, A. (2023). A gamified approach for improving the learning performance of K-6 students using Easter eggs. *Multimedia Tools and Applications*, 1-19.

Tang, J., Hu, X., & Liu, H. (2013). Social recommendation: a review. *Social Network Analysis and Mining*, *3*(4), 1113-1133.

Tejeda-Lorente, Á., Bernabé-Moreno, Julio, H.-Z., Porcel, C., & Herrera-Viedma, E. (2019). A risk-aware fuzzy linguistic knowledge-based recommender system for hedge funds. *Procedia Computer Science*, *162*. https://doi.org/10.1016/j.procs.2019.12.068

Tondello, G. F., Orji, R., & Nacke, L. E. (2017). Recommender systems for personalized gamification. Adjunct publication of the 25th conference on user modeling, adaptation and personalization,

Van Riel, C. B. M., & Fombrun, C. J. (2007). *Essentials of Corporate Communication: Implementing Practices for Effective Reputation Management* (1st ed.). Taylor & Francis. https://doi.org/10.4324/9780203390931

Verma, V., & Aggarwal, R. K. (2020). A comparative analysis of similarity measures akin to the Jaccard index in collaborative recommendations: empirical and theoretical perspective. *Social Network Analysis and Mining*, *10*(1), 1-16.

Villegas, N. M., Sánchez, C., Díaz-Cely, J., & Tamura, G. (2018). Characterizing context-aware recommender systems: A systematic literature review. *Knowledge-Based Systems*, *140*, 173-200. https://doi.org/10.1016/j.knosys.2017.11.003

Xi, N., & Hamari, J. (2019). Does gamification satisfy needs? A study on the relationship between gamification features and intrinsic need satisfaction. *International Journal of Information Management*, *46*, 210-221.

Zhao, Z., Arya, A., Orji, R., & Chan, G. (2020). Effects of a Personalized Fitness Recommender System Using Gamification and Continuous Player Modeling: System Design and Long-Term Validation Study. *JMIR serious games*, *8*(4), e19968.

## Acknowledgments

Not applicable.

## Disclosure statement

No potential conflict of interest was reported by the author(s).

## Funding

There is no funding information applicable to this research.